
\include{def}
\begin{center}
\huge
 The l-State Boson Algebra as a Non-Standard Quantum Double and its Universal
R-Matrix for Yang-Baxer Equation\\

\vspace{0.7cm}
\huge
Wei Li\\
\Large
Departement of Physics,Jilin University,Changchun130023,P.R.China\\
\huge
Chang-Pu Sun\footnote{
\large
Permanent address:Physics Department,Northeast Normal
University,Changchun 130024,P.R.China}\\
\Large
Institute for Teoretical Physics,State University of New York,Stony Brook,
NY 11794-3840 USA\\
\huge
Mo-Lin Ge\\
\Large

Theoretical Physics Division,Nankai Institute of Mathematics,Tianjin 30071
P.R.China\\
\vspace{1cm}
\huge
   Abstract\\

\end{center}

\large

  In this paper we construct a new quantum double by endowing the l-state
bosonalgebra with a non-trivial Hopf algebra structure,which is not a
q-deformation of the Lie algebra or superalgebra.The universal R-matrix
for the Yang-Baxter equation associated with this new quantum group
structure is obtained explicitly.By building the representations of this
 quantum double,we get some R-matrices ,which can result in new
representations of the braid group.

\newpage

\huge
I.Introduction\\
\large
Recently,the quantum Yang-Baxter equation(QYBE) and its quantum group theory
have drawn great interests from both theoretical physicits and mathematician
[1].Drinfeld's quantum double (QD) theory has been proved to be
very powerful in constructing the solutions of  QYBE[2].However,up to now,
most of QD's built explicitly are only those ``q-deformed algebras'' and
``q-deformed Lie groups'',such
as qunatum groups quantum algebras,quantum superalgebras,quantum affine
algebras [3] or
thier paramaterizations[4].The purpose of this paper is to construct a
new quantum double other
than the q-deformed algebras or q-deformed Lie groups.Our study will show that
such a new QD as an
associative algebra contains the usual l-state boson algebra as a subalgebra.
This enables us to endow this boson algebra with a non-cocommutative Hopf
algebraic strucure,which is naturally quasi-triangular.

To proceed our discussion conveniently,we first summarize the central ideas
and the notations in Drinfeld's QD theory.Suppose we are given two Hopf
algebras A ,B and a non-degenerate bilinear form $< , >$:
$A\times B\rightarrow C$(the complex field) satisfying the following
conditions:

1. $$<a, b_1 b_2>=<\Delta_A (a), b_1\otimes b_2>, a\in A,~b_1,~b_2\in B,$$

   $$<a_1 a_2,b>=<a_2\otimes a_1,\Delta_B(b)>,a_1,a_2\in A,b\in A $$

2.$$<1_A,b>=\epsilon_B(b),b\in B,\eqno{(1-1)}$$
  $$<a,1_B>=\epsilon_A(a),a\in A$$
3.$$<S_A(a),S_B(b)>=<a,b>,a\in A,b\in B$$
where for C=A,B,$\Delta_C,\epsilon_C ~and~ S_C $ are the coproduct,counit and
antipode of C respectively;$1_C$ is the unit of C.Drinfeld stated the central
results in the QD theory as follows:
\newtheorem{T1}{Theorem}\begin{T1}There exists a Hopf algrbra D satisfying tyhe
following conditions\\
1.D contains A and B as Hopf subalgebras;\\
2.The mapping $A\times B\rightarrow D:a\otimes b\rightarrow ab$ is an
isomorphism of vector space;\\
3.For any $a\in A,b\in B$,we have multiplication
$$ba=\sum_i <a_i(1),S(b_i(1))><a_i(3),b_i(3)>a_i(2)b_i(2) \eqno{(1-2)}$$
where $c_i(k)(k=1,2,3;c=a,b)$ are defined by
$$\Delta^2(c)=(id\otimes\Delta)\Delta(c)=(\Delta\otimes id)\Delta(c)=
\sum_i c_i(1)\otimes c_i(2)\otimes c_i(3)$$\end{T1}

\newtheorem{T2}{Theorem}\begin{T2}~There exists an unique element

$$\hat{R}=\sum_m a_m\otimes b_m \in A\times B\subset D\times D $$\\
obeying the ``abstract''QYBE\\
$$\hat{R}_{12}\hat{R}_{13}\hat{R}_23=\hat{R}_{23}\hat{R}_{13}\hat{R}_{12},
\eqno{(1-3)}$$
where $a_m$ and $b_m$ are the basis vecors of A and B respectively,and they are
dual
each other;$<a_m,b_n>=\delta_{m,n}$;\\
$$\hat{R}_{12}=\sum_m a_m\otimes b_m\otimes 1,\\
\hat{R}_{13}=\sum_m a_m\otimes 1 \otimes b_m,\\
\hat{R}_{23}=\sum_m 1\otimes a_m\otimes b_m$$
where 1 is the unit of D.\end{T2}

It is pointed out that in the previous QD construction explicitly given
,both subalgebras A and B are non-commutative and non-cocomutative
.But in this paper,our obtained QD contains a commutative but non-
cocommutative subalgebra A and another cocomutatative but non-commutative
subalgebra B .To our best knowledge this is a quite new structure much
different from those built well before.

\vspace{1.4cm}
\huge
II.New quantum double D as a boson algebra\\
\vspace{1cm}
\large

Let consider an associative {(\it {C})} algebra B generated by
{$\bar{N},\bar{a}_1,
\bar{a}_2,...,\bar{a}_l$} satisfying
$$[\bar{N},\bar{a}_i]=\bar{a}_i,[\bar{a}_i,\bar{a}_j]=0,\eqno{(2-1)}$$
We can regard B as a 'half' of the boson algebra generated by creation
operators $\bar{a}_i $,annihilation operators ${a}_i$ and the total number
 operator $\bar{N}$.By endowing B with the following structure operations
$(\Delta,\epsilon,S)$:
\begin{eqnarray}
1.\Delta :&B\rightarrow B\times B:& \Delta(xy)=\Delta(x)\Delta(y) \nonumber \\
&&\Delta(x)=x\otimes 1+1\otimes x \nonumber\\
2.\epsilon:&B\rightarrow \it{C}:&\epsilon (xy)=\epsilon (x)\epsilon
(y)\nonumber\\
&&\epsilon (x)=0,\epsilon(1)=1\nonumber\\
3.S:&B\rightarrow B:&S(xy)=S(y)S(x)\nonumber\\
 &&S(x)=-S(x),S(1)=1\nonumber\\
\end{eqnarray}
for $x=\bar{N},\bar{a}_i$,the algbra B becomes a cocommutative Hopf algebra.
In fact,it is trivial to define such Hopf algebra since we can regard B as
an universal enveloping algebra of a Lie algebra with basis ${\bar{N},
\bar{a}_i(i=1,2,...)}$.However,since B is non-commutative,its dual $B^*=A $
with opposite coproduct $\Delta_A \equiv \Delta $ is non-cocommutative.Now,
we try to determine the generators and the structure relation for A.

According to the PBW theorem,the basis for B can be chosen as
$$\bar{e}(m,n_i)=\bar{e}(m;n_1,n_2,...,n_l)$$
$$=\bar{N}^m \bar{a}_1^{n_1}\bar{a}_2^{n_2}...\bar{a}_l^{n_l}$$
where m,$n_1,n_2,..,n_l\in \it{Z}^+=\{0,1,2,....\}$.On this basis,the
generators N and $a_i(i=1,2,...,l)$ dual to $\bar{N} $and $\bar{a}_i$
respectively can be
defined by the following conditions
$$<\bar{N},N>=1,<x,N>=0$$
$$<\bar{a}_i,a_i>=1,<y,a_i>=o,\eqno{(2-2)} $$
where x is a basis element  other than $\bar {N} $ and y other than $\bar{a}_i
$.

Since the dual of a cocommutative Hopf algebra must be commutative,we
immediately have

{\bf Propositions 1.}{\it A is an Abelian algebra with the commutative
generators N and $a_i$ (i=1,2,...,l).}

Let us extend the bilinear mapping $<  , > $defined only for the generators
to all the vectors pairs in $B\times A.$

{\bf Proposition 2}{\it $$<\bar{N}^m,N^n>=\delta_{m,n}
,<\bar{a}^m_i,a_j^n>= m!\delta_{m,n}\delta_{i,j},\eqno{(2-3)}$$
$$<\bar{e}(m,n_i),N^s
a_1^{r_1}a_2^{r_2}...a_l^{r_l}>=m!n_1!n_2!...n_l!\delta_{m,s}\delta_{n_1,r_1}...\delta_{n_l,r_l},\eqno{(2-4)}
$$}
{\bf Proof}: $$<\bar{N}^m,N^n>=<\Delta(\bar{N}^m),N^{n-1}\otimes N>$$
$$=<\Delta(\bar{N})^m,N^{n-1}\otimes N>$$
$$=\sum_{i=0}^m\frac{m!}{(m-i)!i!}<\bar{N}^i,N^{n-1}><\bar{N}^{m-i},N>$$
$$=m<\bar{N}^{m-1},N^{n-1}>=...=m!\delta_{m,n}$$
The other parts of this proposition can  be proved in a similar way.

As for the Hopf algbra strucure of A,we have to find the coproduct $\Delta$,the
antipode S $\epsilon$ and the counit.\\
{\bf Proposition 3}\\
$$\Delta(N)=N\otimes 1+1\otimes N,\Delta(a_i)=a_i\otimes e^N+1\otimes a_i,i=1,2
,...,l) \eqno{(2-5)}$$
$$S(N)=-N,S(a_i)=-e^{-N}a_i,S(1)=1$$
$$\epsilon(N)=0=\epsilon(a_i),\epsilon(1)=1$$
{\bf Proof.}Notice that the linear form $< ,\Delta(a_i)>$ is non-zero only on
$\bar{a}_i\otimes 1,1\otimes \bar{a}_i,\bar{a}_i\otimes \bar{N}^m (m=1,2,..)$.
In fact,
$$<\bar{a}_i\otimes 1,\Delta(a_i)>=<1\otimes\bar{a_i},\Delta
(a_i)>=<\bar{a_i},a_i>=1$$
$$<\bar{a}_i\otimes \bar{N}^m,\Delta(a_i)>=<\bar{N}^m\bar{a}_i,{a}_i>$$
$$=<\bar{a}_i{(\bar{N}^m+1)}^m,a_i>=1.$$
Consequently,we have
$$\Delta(a_i)=a_i\otimes \sum_{k=0}^{\infty}\frac{N^k}{k!}+1\otimes a_i$$
$$=a_i\otimes e^N+1\otimes a_i.$$
The proof of other parts of this proposition is in the same way.

Having known the Hopf algebraic structure of A,we need to derive the relations
beween A and B so that they are conmbained with each other to form a new
quantum double D.From eqs(1-2) a straightforward,but rather long calculation
results in \\
{\bf Proposition 4}$$[\bar{a}_i,a_j]=\delta _{i,j}(1-e^N),$$
$$[N,a_i]=-a_i,\eqno{(2-6)}$$
$$[N,everything]=0$$

Up to now,from Drinfeld's quantum double theory,we have esteblished new
quasi-triangular Hopf algebra as the quantum double D of A and B,which is
defined by eqs.(2-1),(2-5),(2-6) and the proposition 1.Now,we show how the
l-state
boson algebra can be embeded into the quantum double D as a Hopf subalgebra
and thereby that {\bf the l-state boson algebra possesses  a quantum double
structure!}.To this end,we introduce such an element $E=e^N-1$ that E+1 is
inversable.Then,
$$[a_i,\bar{a}_j]=\delta_{i,j}E,~~~~[E,everything]=0.$$
that is to say,the subalgebra of D genereted by creation operators
$\bar{a}_i$,annihlition operators $a_i$,the number operator $\bar{N}
$ and the unit operator E is just the l-state boson algebra.The strucure
of the quantum double D naturally induces for the l-state boson algebra
a Hopf algebraic construction
$$\Delta(a_i)=a_i\otimes (E+1)+1\otimes a_i,\Delta(E)=E\otimes E
+E\otimes 1+1\otimes E$$
$$S(E)=-\frac{E}{E+1} ,S(a_i)=-\frac{a_i}{E+1} ,S(1)=1,\eqno {(2-7)}$$
$$\epsilon(a_i)=0=\epsilon(E),\epsilon(1)=1$$

\vspace{1cm}
\huge
III.New universal R-matrix for the quantum double D\\
\large

{}From the propositons(1-4),we know that the basis for A dual to
$\bar{e}[m,n_i]$
can be chosen as
$$ e[m.n_i]=\frac{N^ma_1^{n_1}a_2^{n_2}...a_l^{n_l}}{m!n_1!n_2!...n_l!},
\eqno{(3-1)}$$
Then,we immediately write down the universal R-matrix of D
$$\hat{R}=\sum _{[m,n_i]}\bar{e}[m,n_i]\otimes e[m,n_i]$$
$$=exp(\bar{N}\otimes N)\prod _{i=1}^l exp(\bar{a}_i\otimes a_i),\eqno{(3-2)}$$

To get  a finite dimensional R-matrices from eq.(3-2),we have to find the non-
trivial finite dimensional representations of the quantum double D.Here,a
representation T of D in which T(x)=0 for certain generators x of D are thought
to be
trivial.

{\bf Proposition 5}\\
{\it All the non-trivial irreducible representation of D
must be infinite dimensional.}\\
{\bf Proof} Suppose that ther exists a non-trivial finite dimensional
irreducible representation T:D$\rightarrow $End(V).(for simplicity we by x
denote T(x)
as follows).According to the Schrur lemma,the central element N must be a
non-zero scalar $\xi$,ie.,$N=\xi\neq$ 0.For the algebraically-closed
field{\it C},there excists a vector v such that $\bar{N}v=\eta v(\eta\in
{\it C})$.Since a series of eigenvectors $ v,\bar{a}_i,\bar{a}_i^2v,...,
\bar{a}_i^nv$,...,of $\bar{N} $ corresponding to different eigenvalues $
\eta,\eta +1,...,\eta +n$,...,they are independent linearly.Due to the finite
dimension of V there must be a non-zero extreme vector u such that
$$ u=\bar{a}_i^lv,\bar{a}_iu=0$$
Similarly,for other vector series $u,au,a_i^2u,...,$we have $a_i(a_i^{s-1})u
=a^su=o$.It follows from eq.(2-6) that
$$0=\bar{a}_ia^s_iu=(a^s_i\bar{a}_i-sa_i^{s-1}(1-e^N))u=-s(1-e^{\xi})a^{s-1}_iu$$
that is,s=o;the representation space only has dimension zero!

According to the above proposition,the non-trivial finite dimensional
representation of D only is  indecomposable,ie.,reducible but not
completely reducible,if they are not the direct sums of some trivial
representations.Therefore,the non-trivial R-matrices of D should be associated
with
the indecomposable representation of D.

To construct such representations of D explicitly,let us define the Fock-like
space {\it F}(l):
$${\it F}(l)=span\{\mid m_i,p>=\bar{a}_1^{m_1}\bar{a}_2^{m_2}...
\bar{a}_l^{m_l}E^p\mid 0>\mid$$
$$p,m_i=0,1,2,...;i=1,2,...,l\}$$
where the vacuum-like stae $\mid 0>$ obeys
$$a_i\mid 0>=o,\bar{N}\mid 0>=\mu \mid o>,\mu\in{\it C}$$
On this spce,we get an infinite dimensinoal representation
$$\bar{a}_i\mid m_j,p>=\mid m_j+\delta_{i,j},P>$$
$$a_i\mid m_j,p>=m_i\mid m_j-\delta_{i,j},p+1>$$
$$E\mid m-j,p>=\mid m_j,p+1>,\eqno{(3-2)}$$
$$\bar{N}\mid m_j,p>=(m_1+m_2+...+m_l+\mu)\mid m_j,p>$$
where $N=ln(E+1)$.Notice that all the vectors $\mid m_j,p>$ satisfying
$m_1+m_2+...+m_l+p\geq
K\in {\it Z^+} $ span an invariant subspace V(K).Its quotient space
$$Q(K,\mu )={\it F}(l)/V(K)=span\{\mid m^j,p>mod V(K)\mid m_1+m_2+...+m_l+p\leq
K-1\}$$
is finite dimensional and a finite dimensional representation $T^{(\mu,k)}$
of D can be
induced in this quotient space.In next section,we will give explict example
for the representation matrices of D and apply them to construct the matrix
representaton of the universal R-matrix of D.

\vspace{1cm}
\huge
IV.A example of new R-matrix for quantum double D\\
\large

Using the above obstained finite dimensional reepresentation $ T^{(\mu,K)}$
defined by eq.(3-2),the new R-matrix can be constructed as

$$R(H_1,H_2)\doteq R_{\mu_1,\mu_2}^{K_1,K_2}$$
$$=T^{(\mu_1,K_1)}\otimes T^{(\mu_2,K_2)}(\hat{R})\in End[Q(K_1,\mu_1)
\otimes Q(K_2,\mu_2)]$$
The general construction of quantum double theory maintains that the above
R-matrix satisfes the QYBE .Here,the extra parameters $\mu$'s appears
as non-additive dynamic spectrum parameters in physical medels such as
in quantum inverse scattering method and the exactly-solvable models
in statistical mechanics.

For the case of l=2 ,the infinite dimensional representation of D
$$\bar{a}_1\mid m.n,p>=\mid m+1,n,p>$$
$$\bar{a}_2\mid m,n,P>=\mid m.n+1,p>$$
$$a_1\mid m,n,p>=m\mid m-1,n,p+1>$$
$$a_2\mid m,n,p>=n\mid m,n-1,p+1>,\eqno{(4-1)}$$
$$E\mid m,n,p>=\mid m,n,p+1>$$
$$\bar{N}\mid m,n,p>=(m+n+\mu)\mid m,n,p>$$
induces a 4-dimensional representation on $Q(K=2,\mu )$
$$\bar{a}_1=E(2,1),\bar{a}_2=E(3,1),E=E(4,1)$$
$$\bar{N}=\mu E(1,1)+(1+\mu)(E(2,2)+E(3,3))+\mu E(4,4),\eqno{(4-2)}$$
$$a_1=E(4,2),a_2=E(4,3),N=E(4,1)$$
where E(i,j) is $4\times 4 $ matrix unit such that $E(i,j)_{m.n}=
\delta_{im}\delta_{jn}$;$$\mid m,p,n>=\mid m_1=m,m_2=n,p>$$.

Through the universal R-matrix,this explicit representation of D results in
 a 16$\times 16$ R-matrix

$$R=\left[\begin{array}{lccr}
I+\mu N & 0 & 0 & 0 \\
a_1(I+\mu N) & I+(1+\mu)N & 0 & 0 \\
a_2(I+\mu N) & 0 & I+(1+\mu ) & 0 \\
0 & 0 & 0 & I+\mu N
\end{array} \right ]$$

where $\mu$ is a complex parameter and I a $4\times 4$ unit matrix.

\huge
Acknowledgements\\

\large
We wish to express our sincere thanks to Prof.C.N.Yang for drawing our
attentions to the quantum Yang-Baxter equantion and its quantum group theory.
C.P.Sun is supported by Cha Chi Ming fellowship through the CEEC in State
University of New York at Stony Brook.We are also supported in part by the
NFS of China through Northeast Normal University and Nankai Institute of
Mathematics

\newpage

\huge
References\\

\Large
\noindent
1.C.N.Yang,M.L.Ge(eds.){\it Braid Groups,Knot Theory and Statistical
Mechanics},Singapore:World Scientific .1989.\\
2.V.G.Drinfeld,Proc.IMC.Berkeley,1986;P.798\\
N.Y.Reshetikhin,L.A.Taktajian,L.D.Faddeev,Alg.Anal.1(1989),178\\
3.M.Jimbo,Lett.Math.Phys.10(1985)63\\
M.Jimbo,Tipics from representations of $U_q(g)$,in Nankai Lecture Series in
Mathematical Physics,(ed.by M,L.Ge),Singapore:World Scientific,to be
publidhed\\
4.X.F.Liu,C.P.Sun,Science in China A.35(1992)73\\
X.F.Liu,M.L.Ge,Latt.Math.Phys.24(1992),197

\end{document}